\pdfoutput=1
\documentclass[12pt,a4paper]{article}

\usepackage{ifthen} 
\newboolean{pdflatex}
\setboolean{pdflatex}{true} 

\newboolean{articletitles}
\setboolean{articletitles}{true} 

\newboolean{uprightparticles}
\setboolean{uprightparticles}{false} 

\def\paperauthors{LHCb collaboration} 
\def\paperasciititle{Observation of the Lb2JpsiXiK and Xib2JpsiXiPi decays} 
\def\papertitle{Observation of the \LbJpsiXiK and \XibJpsiXiPi decays} 
\def\paperkeywords{{High Energy Physics}, {LHCb}} 
\def\papercopyright{\the\year\ CERN for the benefit of the LHCb collaboration} 
\def\paperlicence{CC BY 4.0 licence}
\def\paperlicenceurl{https://creativecommons.org/licenses/by/4.0/}

\newif\ifEnableSectionTOCLinks
\EnableSectionTOCLinksfalse 

\usepackage[top=1in, bottom=1.25in, left=1in, right=1in]{geometry}

\usepackage{microtype}
\usepackage{lineno}  
\usepackage{xspace} 
\usepackage{caption} 

\usepackage{graphicx}  
\usepackage{color}
\usepackage{colortbl}
\graphicspath{{./figs/}} 

\usepackage{amsmath} 
\usepackage{amssymb}
\usepackage{amsfonts}
\usepackage{upgreek} 

\newcommand*\patchAmsMathEnvironmentForLineno[1]{%
\expandafter\let\csname old#1\expandafter\endcsname\csname #1\endcsname
\expandafter\let\csname oldend#1\expandafter\endcsname\csname
end#1\endcsname
 \renewenvironment{#1}%
   {\linenomath\csname old#1\endcsname}%
   {\csname oldend#1\endcsname\endlinenomath}%
}
\newcommand*\patchBothAmsMathEnvironmentsForLineno[1]{%
  \patchAmsMathEnvironmentForLineno{#1}%
  \patchAmsMathEnvironmentForLineno{#1*}%
}
\AtBeginDocument{%
\patchBothAmsMathEnvironmentsForLineno{equation}%
\patchBothAmsMathEnvironmentsForLineno{align}%
\patchBothAmsMathEnvironmentsForLineno{flalign}%
\patchBothAmsMathEnvironmentsForLineno{alignat}%
\patchBothAmsMathEnvironmentsForLineno{gather}%
\patchBothAmsMathEnvironmentsForLineno{multline}%
\patchBothAmsMathEnvironmentsForLineno{eqnarray}%
}

\usepackage{hyperxmp}

\usepackage[pdftex,
            pdfauthor={\paperauthors},
            pdftitle={\paperasciititle},
            pdfkeywords={\paperkeywords},
            pdfcopyright={Copyright (C) \papercopyright},
            pdflicenseurl={\paperlicenceurl}]{hyperref}

\usepackage[colorinlistoftodos,textsize=scriptsize]{todonotes}

\usepackage[bottom,flushmargin,hang,multiple]{footmisc}

\usepackage[all]{hypcap} 

\usepackage{xspace} 
\usepackage{upgreek}

\def\lhcb   {\mbox{LHCb}\xspace}

\def\MagUp {\mbox{\em Mag\kern -0.05em Up}\xspace}

\ifthenelse{\boolean{uprightparticles}}%
{

 \def\Peta        {\ensuremath{\upeta}\xspace}

 \def\Pmu         {\ensuremath{\upmu}\xspace}

 \def\Ppi         {\ensuremath{\uppi}\xspace}

 \def\Ppsi        {\ensuremath{\uppsi}\xspace}

 \def\PDelta      {\ensuremath{\Delta}\xspace}                 
 \def\PXi         {\ensuremath{\Xi}\xspace}                 
 \def\PLambda     {\ensuremath{\Lambda}\xspace}                 
 \def\PSigma      {\ensuremath{\Sigma}\xspace}                 
 \def\POmega      {\ensuremath{\Omega}\xspace}                 
 \def\PUpsilon    {\ensuremath{\Upsilon}\xspace}
 \let\oldPi\Pi
 \def\PPi         {\ensuremath{\oldPi}\xspace}

 \def\PB      {\ensuremath{\mathrm{B}}\xspace}                 
 \def\PD      {\ensuremath{\mathrm{D}}\xspace}                 
 \def\PJ      {\ensuremath{\mathrm{J}}\xspace}                 
 \def\PK      {\ensuremath{\mathrm{K}}\xspace}                 
 \def\Pb      {\ensuremath{\mathrm{b}}\xspace}                 
 \def\Pc      {\ensuremath{\mathrm{c}}\xspace}

 \def\Pn      {\ensuremath{\mathrm{n}}\xspace}                 
 \def\Pp      {\ensuremath{\mathrm{p}}\xspace}                 

 \def\Ps      {\ensuremath{\mathrm{s}}\xspace}

 \def\thebaroffset{0.0em}
}
{

 \def\Peta        {\ensuremath{\eta}\xspace}

 \def\Pmu         {\ensuremath{\mu}\xspace}

 \def\Ppi         {\ensuremath{\pi}\xspace}

 \def\Ppsi        {\ensuremath{\psi}\xspace}                 
                  
 \mathchardef\PDelta="7101
 \mathchardef\PXi="7104
 \mathchardef\PLambda="7103
 \mathchardef\PSigma="7106
 \mathchardef\POmega="710A
 \mathchardef\PUpsilon="7107
 \mathchardef\PPi="7105
 \def\PB      {\ensuremath{B}\xspace}                 
 \def\PD      {\ensuremath{D}\xspace}                 
 \def\PJ      {\ensuremath{J}\xspace}                 
 \def\PK      {\ensuremath{K}\xspace}                 
 \def\PN      {\ensuremath{N}\xspace}                 
 \def\Pb      {\ensuremath{b}\xspace}                 
 \def\Pc      {\ensuremath{c}\xspace}

 \def\Pn      {\ensuremath{n}\xspace}                 
 \def\Pp      {\ensuremath{p}\xspace}                 

 \def\Ps      {\ensuremath{s}\xspace}

 \def\thebaroffset{0.18em}
}
\newcommand{\offsetoverline}[2][\thebaroffset]{\kern #1\overline{\kern -#1 #2}}%

\makeatletter
\ifcase \@ptsize \relax
  \newcommand{\miniscule}{\@setfontsize\miniscule{4}{5}}
\or
  \newcommand{\miniscule}{\@setfontsize\miniscule{5}{6}}
\or
  \newcommand{\miniscule}{\@setfontsize\miniscule{5}{6}}
\fi
\makeatother

\DeclareRobustCommand{\optbar}[1]{\shortstack{{\miniscule (\rule[.5ex]{1.25em}{.18mm})}
  \\ [-.7ex] $#1$}}

\def\mumu       {{\ensuremath{\Pmu^+\Pmu^-}}\xspace}

\def\squark    {{\ensuremath{\Ps}}\xspace}

\def\cquark    {{\ensuremath{\Pc}}\xspace}

\def\bquark    {{\ensuremath{\Pb}}\xspace}

\def\pion   {{\ensuremath{\Ppi}}\xspace}
\def\piz    {{\ensuremath{\pion^0}}\xspace}
\def\pip    {{\ensuremath{\pion^+}}\xspace}
\def\pim    {{\ensuremath{\pion^-}}\xspace}

\def\kaon    {{\ensuremath{\PK}}\xspace}
\def\Kbar    {{\ensuremath{\offsetoverline{\PK}}}\xspace}
\def\Kb      {{\ensuremath{\Kbar}}\xspace}
\def\KorKbar {\kern \thebaroffset\optbar{\kern -\thebaroffset \PK}{}\xspace}

\def\Kzb     {{\ensuremath{\Kbar{}^0}}\xspace}
\def\Kp      {{\ensuremath{\kaon^+}}\xspace}
\def\Km      {{\ensuremath{\kaon^-}}\xspace}

\def\KS      {{\ensuremath{\kaon^0_{\mathrm{S}}}}\xspace}

\newcommand{\etaz}{\ensuremath{\Peta}\xspace}

\def\D       {{\ensuremath{\PD}}\xspace}

\def\DorDbar {\kern \thebaroffset\optbar{\kern -\thebaroffset \PD}\xspace}

\def\Dp      {{\ensuremath{\D^+}}\xspace}
\def\Dm      {{\ensuremath{\D^-}}\xspace}

\def\DpDm    {\ensuremath{\Dp {\kern -0.16em \Dm}}\xspace}

\def\B       {{\ensuremath{\PB}}\xspace}

\def\BorBbar {\kern \thebaroffset\optbar{\kern -\thebaroffset \PB}\xspace}
\def\Bz      {{\ensuremath{\B^0}}\xspace}

\def\Bd      {{\ensuremath{\B^0}}\xspace}

\def\BdorBdbar {\kern \thebaroffset\optbar{\kern -\thebaroffset \Bd}\xspace}

\def\Bs      {{\ensuremath{\B^0_\squark}}\xspace}

\def\BsorBsbar {\kern \thebaroffset\optbar{\kern -\thebaroffset \Bs}\xspace}

\def\jpsi     {{\ensuremath{{\PJ\mskip -3mu/\mskip -2mu\Ppsi}}}\xspace}
\def\psitwos  {{\ensuremath{\Ppsi{(2S)}}}\xspace}

\def\Y#1S{\ensuremath{\PUpsilon{(#1S)}}\xspace}

\def\proton      {{\ensuremath{\Pp}}\xspace}

\def\neutron     {{\ensuremath{\Pn}}\xspace}

\def\Lz          {{\ensuremath{\PLambda}}\xspace}

\def\LorLbar     {\kern \thebaroffset\optbar{\kern -\thebaroffset \PLambda}\xspace}

\def\Sigmares    {{\ensuremath{\PSigma}}\xspace}

\def\Xires       {{\ensuremath{\PXi}}\xspace}

\def\Xim         {{\ensuremath{\Xires^-}}\xspace}

\def\Lb           {{\ensuremath{\Lz^0_\bquark}}\xspace}

\def\Xib          {{\ensuremath{\Xires_\bquark}}\xspace}
\def\Xibz         {{\ensuremath{\Xires^0_\bquark}}\xspace}
\def\Xibm         {{\ensuremath{\Xires^-_\bquark}}\xspace}

\def\BF         {{\ensuremath{\mathcal{B}}}\xspace}
\def\BR         {\BF}

\newcommand{\decay}[2]{\ensuremath{#1\!\to #2}\xspace} 

\def\to                 {\ensuremath{\rightarrow}\xspace}

\def\AT#1     {\ensuremath{A_{\mathrm{T}}^{#1}}\xspace}           

\def\C#1      {\ensuremath{\mathcal{C}_{#1}}\xspace}                       
\def\Cp#1     {\ensuremath{\mathcal{C}_{#1}^{'}}\xspace}                    
\def\Ceff#1   {\ensuremath{\mathcal{C}_{#1}^{\mathrm{(eff)}}}\xspace}        
\def\Cpeff#1  {\ensuremath{\mathcal{C}_{#1}^{'\mathrm{(eff)}}}\xspace}       
\def\Ope#1    {\ensuremath{\mathcal{O}_{#1}}\xspace}                       
\def\Opep#1   {\ensuremath{\mathcal{O}_{#1}^{'}}\xspace}                    

\newcommand{\nospaceunit}[1]{\ensuremath{\text{#1}}}       
\newcommand{\aunit}[1]{\ensuremath{\text{\,#1}}}       

\newcommand{\tev}{\aunit{Te\kern -0.1em V}\xspace}
\newcommand{\gev}{\aunit{Ge\kern -0.1em V}\xspace}
\newcommand{\mev}{\aunit{Me\kern -0.1em V}\xspace}
\newcommand{\kev}{\aunit{ke\kern -0.1em V}\xspace}
\newcommand{\ev}{\aunit{e\kern -0.1em V}\xspace}
 
\newcommand{\mevc}{\ensuremath{\aunit{Me\kern -0.1em V\!/}c}\xspace}
\newcommand{\gevc}{\ensuremath{\aunit{Ge\kern -0.1em V\!/}c}\xspace}
\newcommand{\mevcc}{\ensuremath{\aunit{Me\kern -0.1em V\!/}c^2}\xspace}
\newcommand{\gevcc}{\ensuremath{\aunit{Ge\kern -0.1em V\!/}c^2}\xspace}

\def\mum  {\ensuremath{\,\upmu\nospaceunit{m}}\xspace}

\def\fb   {\ensuremath{\aunit{fb}}\xspace}
\def\invfb   {\ensuremath{\fb^{-1}}\xspace}

\def\ps   {\ensuremath{\aunit{ps}}\xspace}

\newcommand{\chisq}{\ensuremath{\chi^2}\xspace}

\def\gsim{{~\raise.15em\hbox{$>$}\kern-.85em
          \lower.35em\hbox{$\sim$}~}\xspace}
\def\lsim{{~\raise.15em\hbox{$<$}\kern-.85em
          \lower.35em\hbox{$\sim$}~}\xspace}

\def\pt         {\ensuremath{p_{\mathrm{T}}}\xspace}

\def\ptot       {\ensuremath{p}\xspace}

\def\evtgen     {\mbox{\textsc{EvtGen}}\xspace}

\def\geant      {\mbox{\textsc{Geant4}}\xspace}

\def\photos     {\mbox{\textsc{Photos}}\xspace}

\def\pythia     {\mbox{\textsc{Pythia}}\xspace}

\def\tell1  {TELL1\xspace}
\def\ukl1   {UKL1\xspace}

\newcommand{\ie}{\mbox{\itshape i.e.}\xspace}

\newcommand{\lhcborcid}[1]{\href{https://orcid.org/#1}{\hspace*{0.1em}\raisebox{-0.45ex}{\includegraphics[width=1em]{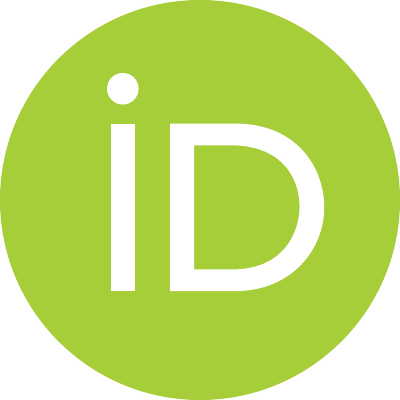}}}}

\hypersetup{
  colorlinks   = true, 
  urlcolor     = blue, 
  linkcolor    = blue, 
  citecolor    = red   
}

\ifEnableSectionTOCLinks
    \usepackage[explicit]{titlesec} 
    
    \let\oldcontentsline\contentsline
    \renewcommand

    \titleformat{\section}{\normalfont\Large\bf}{\hyperlink{tocsection.\thesection}{{\thesection} \parbox[t]{\dimexpr\textwidth-1pc}{#1}}}{1pc}{}

    \titleformat{\subsection}{\normalfont\bf}{\hyperlink{tocsubsection.\thesubsection}{{\thesubsection} \parbox[t]{\dimexpr\textwidth-1pc}{#1}}}{1pc}{}

    \titleformat{name=\section,numberless}[display]{}{}{0pt}{\normalfont\Huge\bfseries #1}
\fi

\usepackage{cite} 
\usepackage{mciteplus}

\def\LbJpsiXiK     {\ensuremath{\mbox{\decay{\Lb}{\jpsi \Xim \Kp}}}\xspace}
\def\LbJpsiL       {\ensuremath{\mbox{{\decay{\Lb}{\jpsi \Lz}}}}\xspace}
\def\XibJpsiXiPi   {\ensuremath{\mbox{{\decay{\Xibz}{\jpsi \Xim \pip}}}}\xspace}
\def\XibJpsiXi     {\ensuremath{\mbox{{\decay{\Xibm}{\jpsi \Xim}}}}\xspace}
\def\BdJpsiKS      {\ensuremath{\mbox{{\decay{\Bz}{\jpsi \KS}}}}\xspace}

\usepackage{relsize}

\def\BRRatLb       {\ensuremath{{\frac{\mathlarger{\BR}(\LbJpsiXiK)}{\mathlarger{\BR}(\LbJpsiL)}}}\xspace}
\def\BRRatXib      {\ensuremath{{\frac{\mathlarger{\BR}(\XibJpsiXiPi)}{\mathlarger{\BR}(\XibJpsiXi)}}}\xspace}

\def\BRRatLbText       {\ensuremath{{\BR(\LbJpsiXiK)/\BR(\LbJpsiL)}}\xspace}
\def\BRRatXibText      {\ensuremath{{\BR(\XibJpsiXiPi)/\BR(\XibJpsiXi)}}\xspace}

\begin{document}

\renewcommand{\thefootnote}{\fnsymbol{footnote}}
\setcounter{footnote}{1}


\begin{titlepage}
\pagenumbering{roman}

\vspace*{-1.5cm}
\centerline{\large EUROPEAN ORGANIZATION FOR NUCLEAR RESEARCH (CERN)}
\vspace*{1.5cm}
\noindent
\begin{tabular*}{\linewidth}{lc@{\extracolsep{\fill}}r@{\extracolsep{0pt}}}
\ifthenelse{\boolean{pdflatex}}
{\vspace*{-1.5cm}\mbox{\!\!\!\includegraphics[width=.14\textwidth]{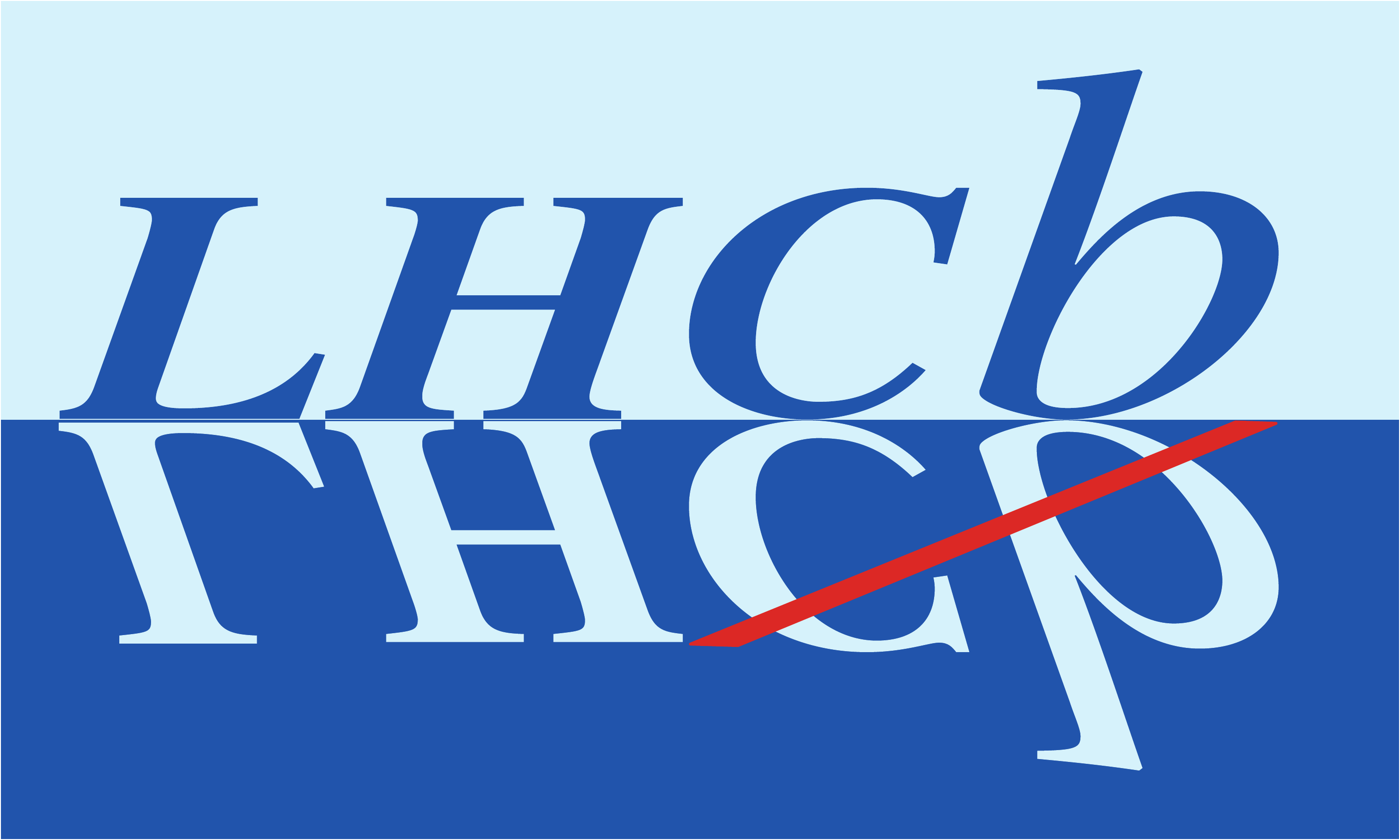}} & &}%
{\vspace*{-1.2cm}\mbox{\!\!\!\includegraphics[width=.12\textwidth]{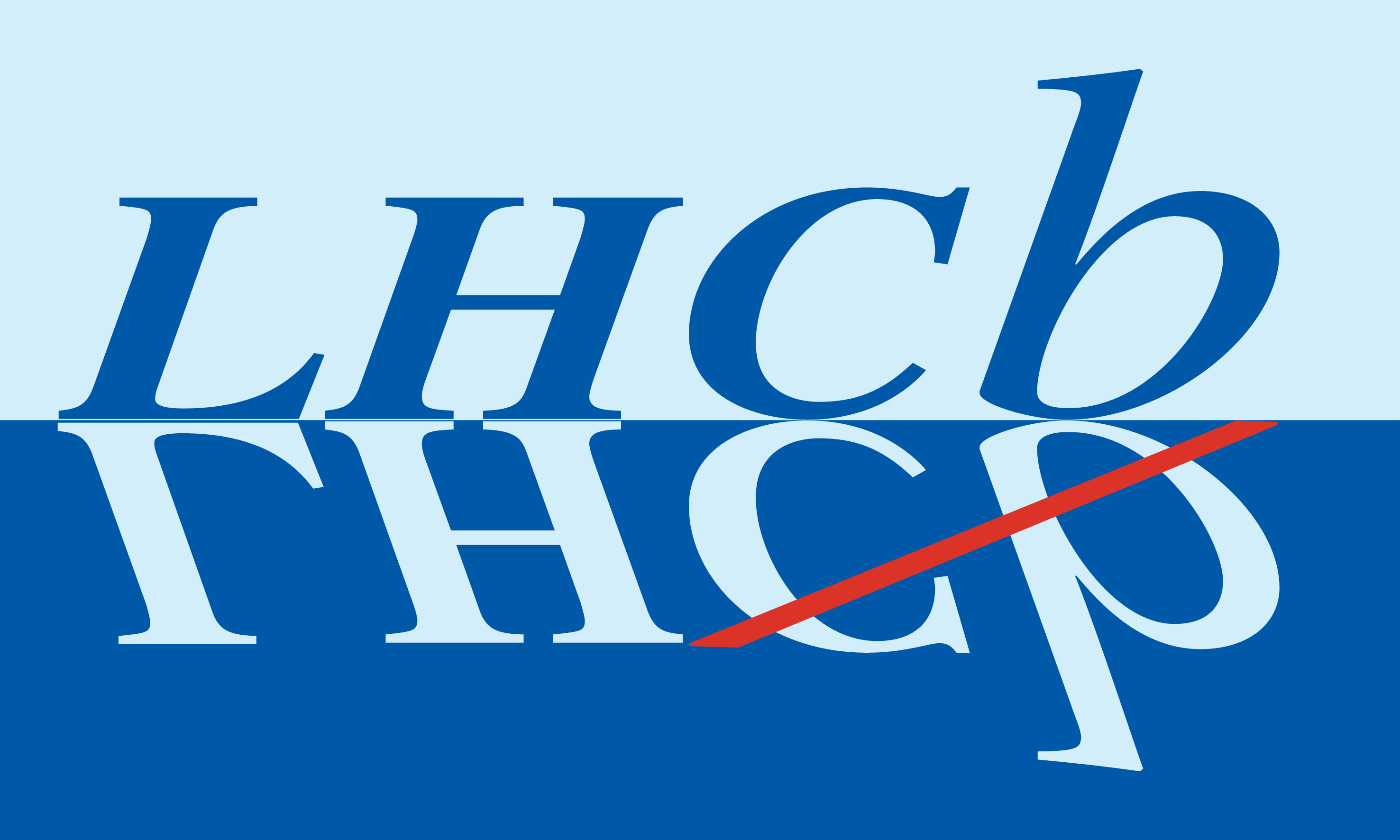}} & &}%
\\
 & & CERN-EP-2024-337 \\  
 & & LHCb-PAPER-2024-049 \\  
 & & September 8, 2025  \\ 
 & & \\
\end{tabular*}

\vspace*{3.0cm}

{\normalfont\bfseries\boldmath\huge
\begin{center}
  \papertitle 
\end{center}
}

\vspace*{2.0cm}

\begin{center}
\paperauthors\footnote{Authors are listed at the end of this paper.}
\end{center}

\vspace{\fill}

\begin{abstract}
  \noindent
  The first observation of the \XibJpsiXiPi decay and the most precise measurement of the branching fraction of the \LbJpsiXiK decay are reported, using proton-proton collision data from the LHCb experiment collected in 2016--2018 at a centre-of-mass energy of 13\tev, corresponding to an integrated luminosity of 5.4\invfb. Using the \LbJpsiL and \XibJpsiXi decays as normalisation channels, the ratios of branching fractions are measured to be
\begin{align*}
    \BRRatLb &= (1.17 \pm 0.14 \pm 0.08)\times 10^{-2} \, ,\\
    \BRRatXib & = (11.9 \pm 1.4 \pm 0.6)\times 10^{-2}\, ,
\end{align*}
  where the first uncertainty is statistical and the second systematic. 

\end{abstract}

\vspace*{2.0cm}

\begin{center}
  Published in The European Physical Journal C 85 (2025) 812
\end{center}

\vspace{\fill}

{\footnotesize 
\centerline{\copyright~\papercopyright. \href{\paperlicenceurl}{\paperlicence}.}}
\vspace*{2mm}

\end{titlepage}


\newpage
\setcounter{page}{2}
\mbox{~}
%
%
%
%

\renewcommand{\thefootnote}{\arabic{footnote}}
\setcounter{footnote}{0}


\cleardoublepage


\pagestyle{plain} 
\setcounter{page}{1}
\pagenumbering{arabic}


\section{Introduction}

The existence of pentaquark states has been predicted since the establishment of the quark model~\cite{GellMann:1964nj,*Zweig:352337}.
In 2015 the first pentaquark states were observed by the \lhcb experiment in the spectrum of \decay{\Lb}{\jpsi\proton\Km} decays~\cite{LHCb-PAPER-2015-029}.\footnote{Charge conjugation is implied throughout this paper.} This observation sparked significant interest in the community,  leading to various interpretations on the internal structure of these states~\cite{PhysRevLett.124.072001,PhysRevD.92.094003,PhysRevD.103.L111503, Eides2018, Santopinto_2017}. In the subsequent years, numerous pentaquark states have been discovered in \bquark-hadron decays to final states containing charmonium~\cite{LHCb-PAPER-2019-014, LHCb-PAPER-2020-039, LHCb-PAPER-2022-031, LHCb-PAPER-2021-018}. 
After the observation of pentaquarks with strangeness $S=0$ and $S=1$ in the $\jpsi\proton$ and $\jpsi\Lz$ systems, the search for a pentaquark with $S=-2$ in the $\jpsi\Xim$ system is the next natural step~\cite{PhysRevD.103.054025,Azizi2022}.
In this paper, the foundation for these searches is laid by the observation of the \LbJpsiXiK and \XibJpsiXiPi decays.

\begin{figure}[!b]
    \centering
    \includegraphics[page=2,width=0.48\linewidth]{figs/Fig1.pdf}
    \includegraphics[page=1,width=0.48\linewidth]{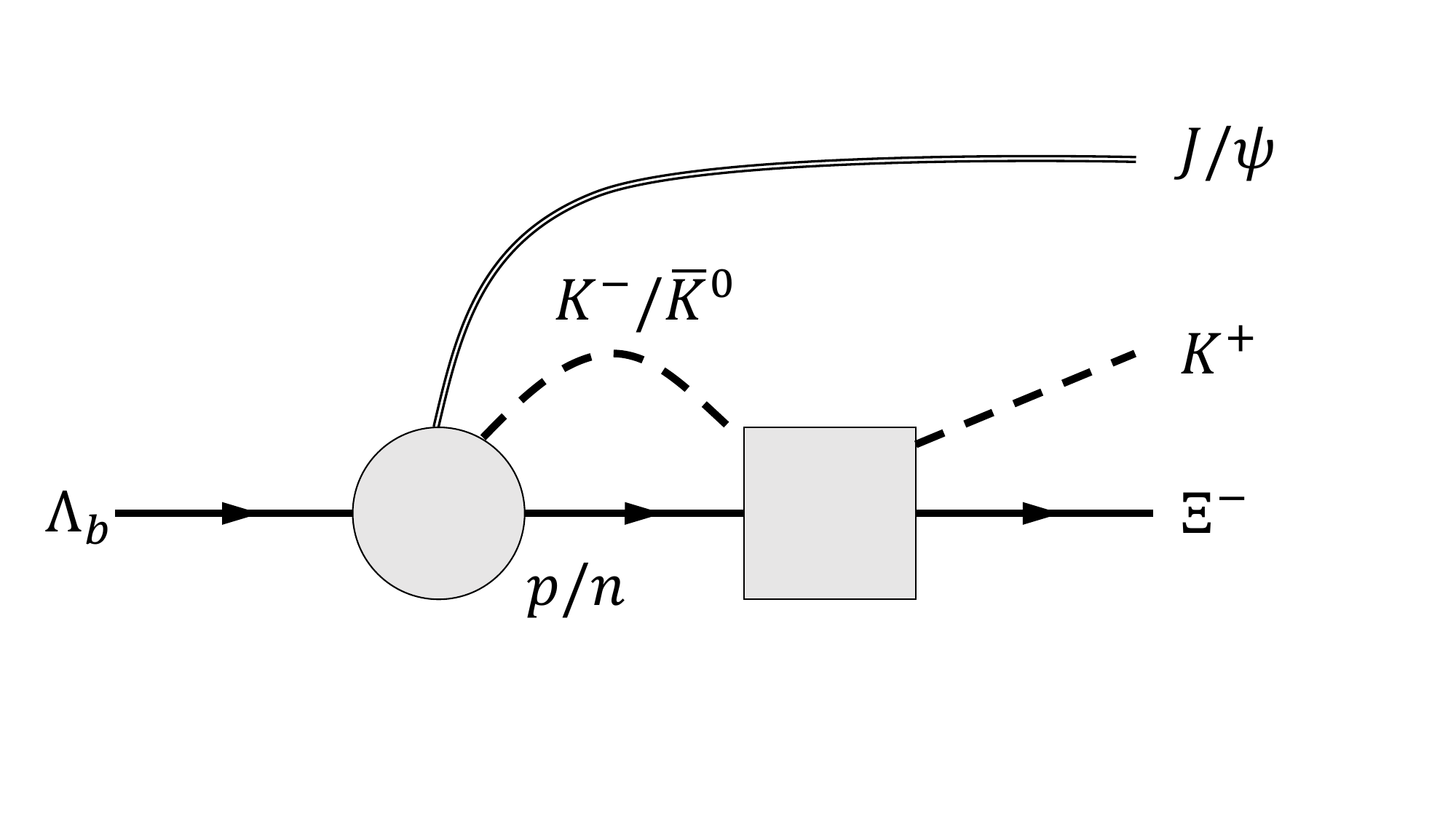}
    \caption{(Left) Production of a $\proton\Km$, $\neutron\Kzb$ or $\Lz\etaz$ pair from the weak decay \LbJpsiL via a hadronisation mechanism.  (Right) Final-state interaction of the meson-baryon pair, where the double line denotes the \jpsi meson, solid lines represent baryons and dashed lines denote the pseudoscalar mesons. The shaded circle and square stand for the production mechanism of the $\jpsi\proton\Km$ or $\jpsi\neutron\Kzb$, and the meson-baryon scattering matrix, respectively.
    }
    \label{fig:fsi}
\end{figure}

Besides the search for exotic states in decay products including a charmonium resonance, the studied decay channels give insights into the dynamics  of baryon-meson systems with one or two units of strangeness~\cite{Strangeness}.
Of particular interest are $\Kb\PN$ meson-baryon interactions which are used to describe the $\Lz(1405)$ resonance using Unitarised Chiral Perturbation Theory~\cite{Cabrera_2016}. These interactions have been studied using data from ${\decay{\Km\proton}{\kaon\PN, \pion\Sigmares, \piz\Lz}}$ inelastic scattering, near threshold production of the $\Lz(1405)$ resonance~\cite{PhysRevC.92.015206}. Scattering studies at higher energies --- above the \kaon\Xires threshold --- are interesting since they are sensitive to next-to-leading order terms of the Chiral Lagrangian~\cite{PhysRevC.92.015206}. Most of the \decay{\Kb\PN}{\kaon\Xires} data come from antikaon-proton scattering, which includes both isospin $I=0$ and $I=1$ contributions. 
In contrast, \Km\proton scattering, which occurs in the decay process \LbJpsiXiK, is restricted to $I=0$ contribution in order to preserve the overall isospin of the decay. Studying this weak process therefore allows the $I=0$ component to be isolated, thus enhancing our understanding of meson-baryon interactions~\cite{Roca2015}. 
In the \LbJpsiXiK decay, the direct production of the \Xim\Kp final state is forbidden in the quark model, and is only reached via a two-step process, as shown in Fig.~\ref{fig:fsi}. First, the \decay{\Lb}{\jpsi\{\Lz\etaz, \proton\Km, \neutron\Kb\}} decay proceeds via the weak interaction. Second, the meson-baryon pair produced in the decay scatters into a \Xim\Kp pair~\cite{PhysRevC.92.015206}.
A measurement of the $m(\Xim\Kp)$ spectrum in the \LbJpsiXiK decay isolates the $I=0$ contribution, and allows the $I=0$ and $I=1$ contributions in previous data~\cite{PhysRevC.92.015206} to be disentangled. As a consequence, the modelling of meson-baryon interaction above the $\Lz(1405)$ resonance can be improved.

The \XibJpsiXiPi decay has not yet been observed. 
This decay is expected to receive contributions from $\decay{\Xibz}{\jpsi\Xires^*(\Xim\pip)}$ processes, where $\Xires^*$ are excited states of the \Xires baryon. The first excitation of the \Xires baryon, the $\Xires(1530)^0$ state, is flavour-symmetric and would be suppressed in the $\decay{\Xibz}{\Xires(1530)^0}$ transition due to the \Xibz state being flavour-antisymmetric~\cite{XiRes1530}. Measuring the $\decay{\Xibz}{\jpsi\Xires(1530)^0}$ decay rate is a method to quantify the mixing of symmetric and antisymmetric flavour states.
The \Xim\pip invariant-mass spectrum in \XibJpsiXiPi decays can help explore possible mixing between flavour-symmetric and -antisymmetric contributions.
Furthermore, the \XibJpsiXiPi channel provides an excellent opportunity to search for charmonium pentaquark states.
The first step towards the study of the decay dynamics  is to observe these decay channels and analyse their mass spectra. This paper details the first measurement of the \XibJpsiXiPi branching fraction and improves the precision of the \LbJpsiXiK branching fraction with respect to the latest measurement by the CMS collaboration~\cite{Hayrapetyan:2887755}. The measurements reported in this paper use \proton\proton collision data collected by the \lhcb experiment in 2016--2018 at a centre-of-mass energy of 13\tev, corresponding to an integrated luminosity of 5.4\invfb.

The \LbJpsiXiK and \XibJpsiXiPi signal channels are reconstructed through the \decay{\jpsi}{\mumu}, \decay{\Xim}{\Lz\pim} and \decay{\Lz}{\proton\pim} decays. The \LbJpsiL and \XibJpsiXi decay modes are used as normalisation channels due to their similar experimental signature. The choice of normalisation channels enables the cancellation of the dependence on \bquark-hadron production rates and minimises systematic uncertainties related to final-state particle reconstruction.
The ratios of branching fractions are measured as
\begin{align}
    \BRRatLb=&\frac{N(\LbJpsiXiK)}{N(\LbJpsiL)} \cdot \frac{\varepsilon(\LbJpsiL)}{\varepsilon(\LbJpsiXiK)}\cdot \BF(\decay{\Xim}{\Lz\pim})\, , \label{eq:BFRatioLb}
    \\
    \BRRatXib=&\frac{N(\XibJpsiXiPi)}{N(\XibJpsiXi)} \cdot \frac{\varepsilon(\XibJpsiXi)}{\varepsilon(\XibJpsiXiPi)}\, ,
    \label{eq:BFRatioXib}
\end{align}
where \BF is the branching fraction, $N$ is the number of observed decays and $\varepsilon$ is the total efficiency for the corresponding decay. 
An event selection is developed using simulated samples to optimise the search sensitivity. Correct estimation of the efficiencies relies on the accurate modelling of the physics processes and detector response in the simulation. The normalisation channels are used to correct for inaccuracies in the simulation for observables with a significant disagreement between simulation and data.
After the selection is completed, the signal and normalisation yields are obtained from fits to the invariant-mass distributions of candidates in the selected data. In addition, the background-subtracted invariant-mass distributions of the $\Xim\Kp$, $\jpsi\Kp$ and $\jpsi\Xim$ pairs for the \LbJpsiXiK channel, and the $\Xim\pip$, $\jpsi\pip$ and $\jpsi\Xim$ pairs for the \XibJpsiXiPi channel are investigated. The background-subtracted samples are obtained using the \texttt{sPlot} technique~\cite{Pivk:2004ty, DEMBINSKI2022167270}, a statistical method that separates signal and background distributions using event-by-event weights derived from a fit to discriminating variables. The \bquark-baryon invariant-mass is chosen as discriminant variable.
The invariant-mass signal regions were not examined until the full procedure had been finalised. 


\section{Detector and simulation}
\label{sec:Detector}
The \lhcb detector~\cite{LHCb-DP-2008-001,LHCb-DP-2014-002} is a single-arm forward spectrometer covering the \mbox{pseudorapidity} range $2<\eta <5$, designed for the study of particles containing \bquark or \cquark quarks. The detector used to collect data in 2015--2018 (Run 2) includes a high-precision tracking system consisting of a silicon-strip vertex detector (VELO) surrounding the $pp$ interaction region, a large-area silicon-strip detector, the Tracker Turicensis (TT), located upstream of a dipole magnet with a bending power of about $4{\mathrm{\,T\,m}}$, and three stations of silicon-strip detectors, straw drift tubes (T-stations) placed downstream of the magnet.
The tracking system provides a measurement of the momentum, \ptot, of charged particles with a relative uncertainty that varies from 0.5\% at low momentum to 1.0\% at 200\gevc.
The minimum distance of a track to a primary $pp$ collision vertex (PV), the impact parameter (IP), is measured with a resolution of $(15+29/\pt)\mum$, where \pt is the component of the momentum transverse to the beam, in\,\gevc.
Different types of charged hadrons are distinguished using information from two ring-imaging Cherenkov detectors. 
Muons are identified by a system composed of alternating layers of iron and multiwire proportional chambers.
The online event selection~\cite{LHCb-DP-2019-001} is performed by a trigger which consists of a hardware stage, based on information from the calorimeter and muon systems, followed by two software stages, which apply a partial and full event reconstruction, respectively.
At the hardware trigger stage, events are required to have muon or dimuon candidates with high \pt. The software trigger requires a two-track secondary vertex with a significant displacement from any PV. At least one charged particle must have $\pt > 1.6\gevc$ and be inconsistent with originating from a PV.

Simulated samples are used to develop the selection strategy, compute the efficiencies and model the shape of the invariant-mass distribution of the signal decays. 
In the simulation, $pp$ collisions are generated using \pythia~\cite{Sjostrand:2006za} with a specific \lhcb configuration~\cite{LHCb-PROC-2010-056}. Decays of unstable particles are described by \evtgen~\cite{Lange:2001uf}, in which final-state radiation is generated using \photos~\cite{davidson2015photos}. The interaction of the generated particles with the detector, and its response, are implemented using the \geant toolkit~\cite{Allison:2006ve, *Agostinelli:2002hh} as described in Ref.~\cite{LHCb-PROC-2011-006}. For each generated signal decay the underlying \proton\proton interaction is reused multiple times~\cite{LHCb-DP-2018-004}.


\section{Candidate selection and efficiencies}
\label{sec:sel}

The selection begins by identifying well-reconstructed tracks that do not originate from any PV. 
Candidates with two oppositely charged tracks identified as muons are selected. Each muon candidate must fulfil $\pt>250\mevc$ requirement. Dimuon pairs originating from a common vertex that is significantly displaced from any PV are combined to form \jpsi candidates. The dimuon pair must have a reconstructed invariant mass compatible with the nominal mass of the \jpsi meson.  
Oppositely charged tracks assigned to proton and pion mass hypotheses are combined to form \Lz candidates. The \proton\pim invariant mass is required to be within $\pm10\mevcc$ of the known \Lz mass~\cite{PDG2024}. The \Lz candidates are further combined with another particle, assigned the pion mass hypothesis, to reconstruct \Xim candidates. The selected \Lz\pim pairs must have a \pt larger than 250\mevcc and an invariant mass within $\pm20\mevcc$ of the known \Xim mass~\cite{PDG2024}. For reference, the mass resolution of the \Lz and \Xim signals are about 1.7\mevcc and 3.0\mevcc, respectively.
Given the relatively large lifetime of the \Lz and \Xim baryons, most of these candidates decay downstream of the VELO, but before reaching the TT. Tracks that are reconstructed only from hits in the TT and the T-stations are called downstream tracks. Tracks that are reconstructed using hits in the VELO, the T-stations, and optionally in the TT, are called long tracks.
Long tracks profit from the superior resolution of the vertex detector and have better mass, momentum and vertex position resolution than downstream tracks.
The considered \decay{\Lz}{\proton\pim} candidates can be reconstructed from two long tracks or two downstream tracks.
 Similarly, the \decay{\Xim}{\Lz\pim} decay is reconstructed using three main categories of track-type combinations: (i) both \Lz and \pim candidates reconstructed only from long tracks, (ii) both \Lz and \pim candidates reconstructed only from downstream tracks, and (iii) the \Lz candidate reconstructed from two downstream tracks and the accompanying \pim candidate reconstructed as a long track. 
The \Lb (\Xibz) signal candidate is formed by combining the \jpsi and \Xim candidates with a long track identified as a \Kp (\pip) candidate with $\pt>250\mevc$. Similarly, \jpsi and \Lz (\Xim) candidates are combined to form \Lb (\Xibm) candidates corresponding to the normalisation channel decay. Both \Lz and \Xim candidates are required to have a decay time greater than 2\ps to reject background from short-lived particles. The \Lb (\Xib) decay vertices must have a good fit quality and be significantly displaced from any PV. 
A kinematic fit~\cite{Hulsbergen:2005pu} is performed to improve the \bquark-hadron mass resolution by constraining the mass hypothesis of the \jpsi, \Xim and \Lz candidates to their known masses~\cite{PDG2024} and requiring the \bquark hadron to originate from its associated PV\footnote{The associated PV is defined as the PV that fits best to the flight direction of the \bquark-hadron candidate.}. Candidates with a poor quality of the kinematic fit or a \Lb (\Xib) invariant mass from the kinematic fit outside the range $[5360,5900]$ ($[5700,6100]$)\mevcc are rejected.

The final step of the selection for the four decays channels is to apply a boosted decision tree~(BDT)~\cite{Breiman,AdaBoost}  classifier implemented using the XGBoost toolkit~\cite{10.1145/2939672.2939785}. Each classifier is trained to distinguish between the specific \bquark-hadron decay and combinatorial background. Simulation samples are used as a proxy for the specific \bquark-hadron decays. The background samples are taken from the high invariant-mass sideband [5700, 5900] ([5900, 6100])\mevcc of selected \Lb (\Xib) data candidates. The BDT classifier exploits the discriminating power of the IP of the \pim meson from the \Xim decay, and \Lz decay products with respect to any PV; the difference in the vertex-fit \chisq of a given PV reconstructed with and
without the \bquark-hadron and the muons; the \chisq of the kinematic fit; the decay time of the \bquark hadron and \Xim baryon; and the \pt of the \Xim candidate. Most of the training variables are common between the signal and normalisation channels, with the exception of variables associated with the \Xim candidate that are excluded when analysing the \LbJpsiL mode. The BDT working point is chosen by maximising a figure of merit~\cite{Punzi:2003bu}, defined as $\varepsilon_\text{s}/(\sqrt{B} + 2.5)$, where $\varepsilon_\text{s}$ is the efficiency of the requirement on the BDT output for simulated signal events, and $B$ is the background yield in a region within 3 sigmas around the nominal \bquark-hadron mass, extrapolated from the high invariant-mass sideband. The BDT classifier removes more than 99\% of the background while retaining approximately 70\% of the signal in the four considered decay modes. 
After the described selection, more than one candidates are selected per \proton\proton in less than 1\% of the events. In this cases, the candidate with the highest BDT score is kept.

The total efficiency is calculated as the product of efficiencies of detector acceptance, reconstruction, and selection.
Accurate simulation is crucial to determine such selection efficiencies, and also to train the BDT algorithms and model the signal mass shapes. Discrepancies between data and simulation are observed and corrected for in the \bquark-baryon \pt, the \chisq of the kinematic fit, angular and two-body invariant-mass distributions, and particle identification and tracking efficiencies. The corrections are derived sequentially, \ie adjusting  the distribution for one variable at a time. Potential correlations between variables are accounted for by deriving each correction factor on top of previous ones.
The particle identification efficiencies for the different particle species are measured using charm hadron data samples reconstructed without the use of PID information~\cite{LHCb-DP-2018-001}. The correction factors for the tracking efficiencies are derived from a dedicated data sample of \decay{\jpsi}{\mumu} decays~\cite{LHCb-DP-2013-002}. 
The correction factors for \bquark-baryon \pt and the \chisq of the kinematic fit are extracted from a comparison of \LbJpsiL and \XibJpsiXi decays between background-subtracted data and simulated samples and applied to both signal and normalisation channels. The angular distribution could differ between the signal and normalisation channels, and thus the above correction factors are derived independently for each of the four channels.
The decay dynamics of the signal channels are also corrected for the $m(\jpsi\Xim)$ and $m(\Xim\Kp)$, or $m(\Xim\pip)$, distributions. The variables corrected are those that demonstrated a significant correlation with the efficiency. These correction factors are derived from the \LbJpsiXiK and \XibJpsiXiPi signal channels.
No significant correlation is observed between the discriminant variable used for the background-subtraction method, the \bquark-hadron invariant mass improved by the kinematic fit, and the corrected variables. After applying the correction, a good agreement is observed between the background-subtracted data and the simulation on all variables used for the selection, including the BDT response.
The ratios of efficiencies between the \Lb and \Xib modes, required in Eqs.~\ref{eq:BFRatioLb} and \ref{eq:BFRatioXib}, are determined to be
\begin{align*}
    \frac{\varepsilon(\LbJpsiL)}{\varepsilon(\LbJpsiXiK)} = 5.30\pm0.05\, ,\\
    \frac{\varepsilon(\XibJpsiXi)}{\varepsilon(\XibJpsiXiPi)} = 1.64\pm0.02 \, ,
\end{align*}
where the uncertainty is due to the limited size of the simulation samples. This uncertainty is propagated to the branching fraction ratios as a source of systematic uncertainty in Sec.~\ref{sec:systematics}.

\section{Signal yield determination}
\label{sec:massfit}

An extended unbinned maximum-likelihood fit is performed to the \bquark-hadron invariant-mass spectrum of the \LbJpsiXiK and \LbJpsiL modes in the range \mbox{[5360, 5900]\mevcc}. An analogous, independent fit is performed to the \XibJpsiXiPi and \XibJpsiXi candidates in the range \mbox{[5700, 6100]\mevcc}. 
The invariant-mass shape of the \LbJpsiL decay mode is modelled by a Hypatia function~\cite{Santos:2013gra}.
The signal for the other three channels is modelled by the sum of two Crystal Ball functions~\cite{Skwarnicki:1986xj} with common mean and width.
The tail parameters for both signal models, Hypatia and two Crystal Ball functions, are fixed according to simulation.
The shape of the invariant-mass distribution of the combinatorial background is modelled by an exponential function with a slope varying freely in the fit.
Non-negligible contribution from decays of other \bquark-hadrons  is only observed in the \LbJpsiL channel. The two main background sources are \BdJpsiKS and \XibJpsiXi decays.
The \BdJpsiKS decay can be misidentified as a \LbJpsiL candidate if the proton hypothesis is wrongly assigned to the \pip meson in the \KS decay. Additionally, missing the pion from the \decay{\Xim}{\Lz\pim} decay leads to a \XibJpsiXi decay that is partially reconstructed as a \LbJpsiL candidate.
The invariant-mass shapes of the misidentified \BdJpsiKS and partially reconstructed \XibJpsiXi backgrounds are derived from simulation in terms of a nonparametric probability density function using a kernel density estimation method~\cite{roofit,CRANMER2001198} and a Johnson $S_{U}$ function~\cite{JohnsonSU}, respectively. The shapes of these functions are fixed from simulation.

\begin{figure}[tb]
    \centering
    \includegraphics[width=0.45 \textwidth]{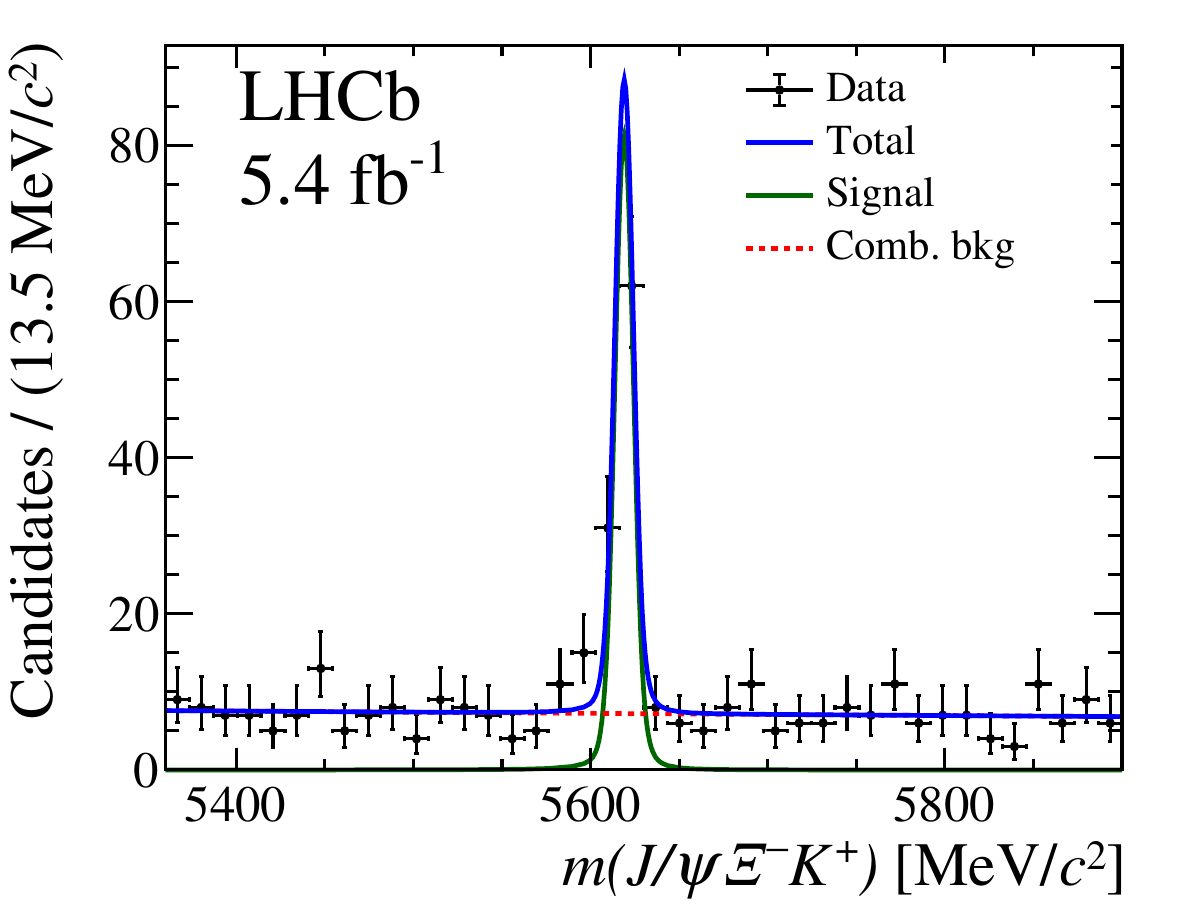}
    \includegraphics[width=0.45 \textwidth]{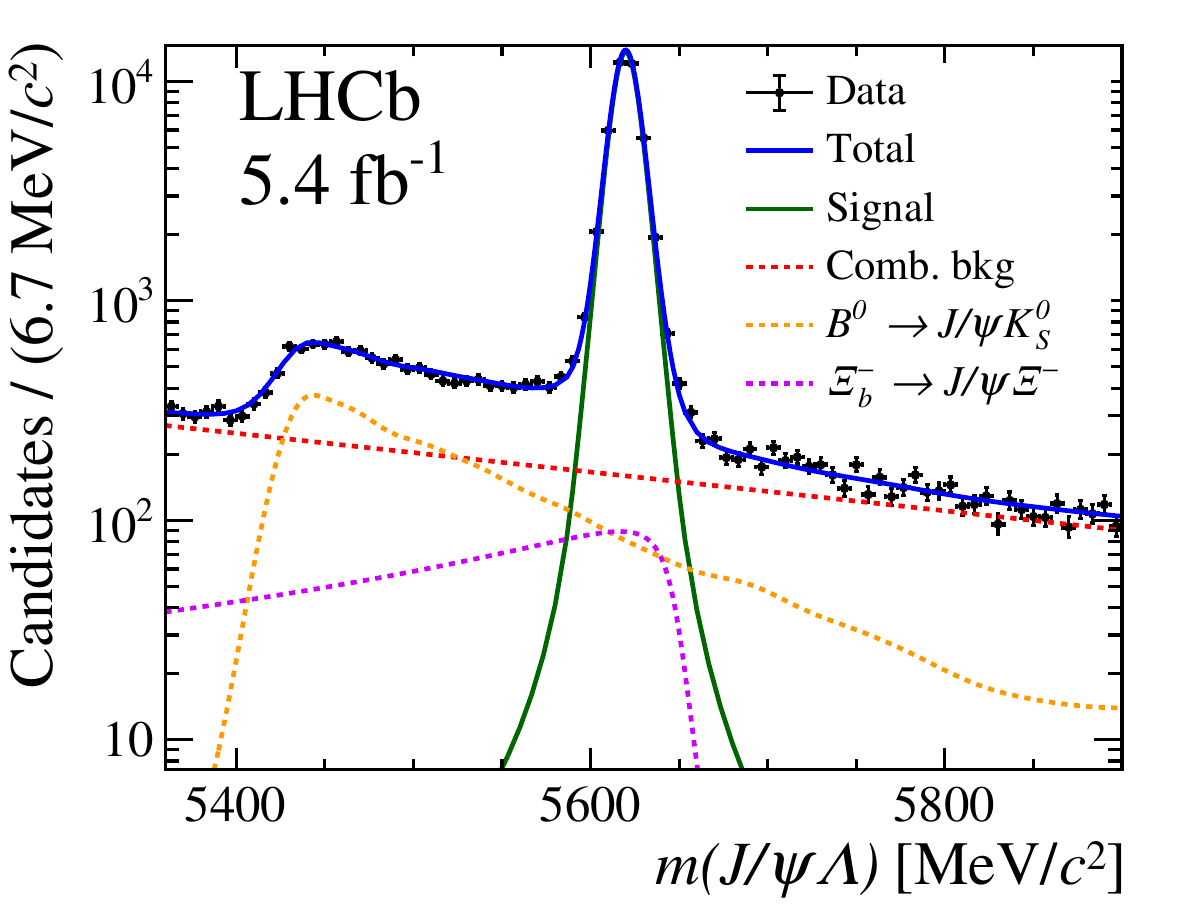}
    \caption{Invariant-mass distribution of selected (left) \LbJpsiXiK and (right) \LbJpsiL  candidates with the result of the fit also shown. 
    }
    \label{fig:Massfit_data_Lb}
\end{figure}

\begin{figure}[tb]
    \centering
    \includegraphics[width=0.45 \textwidth]{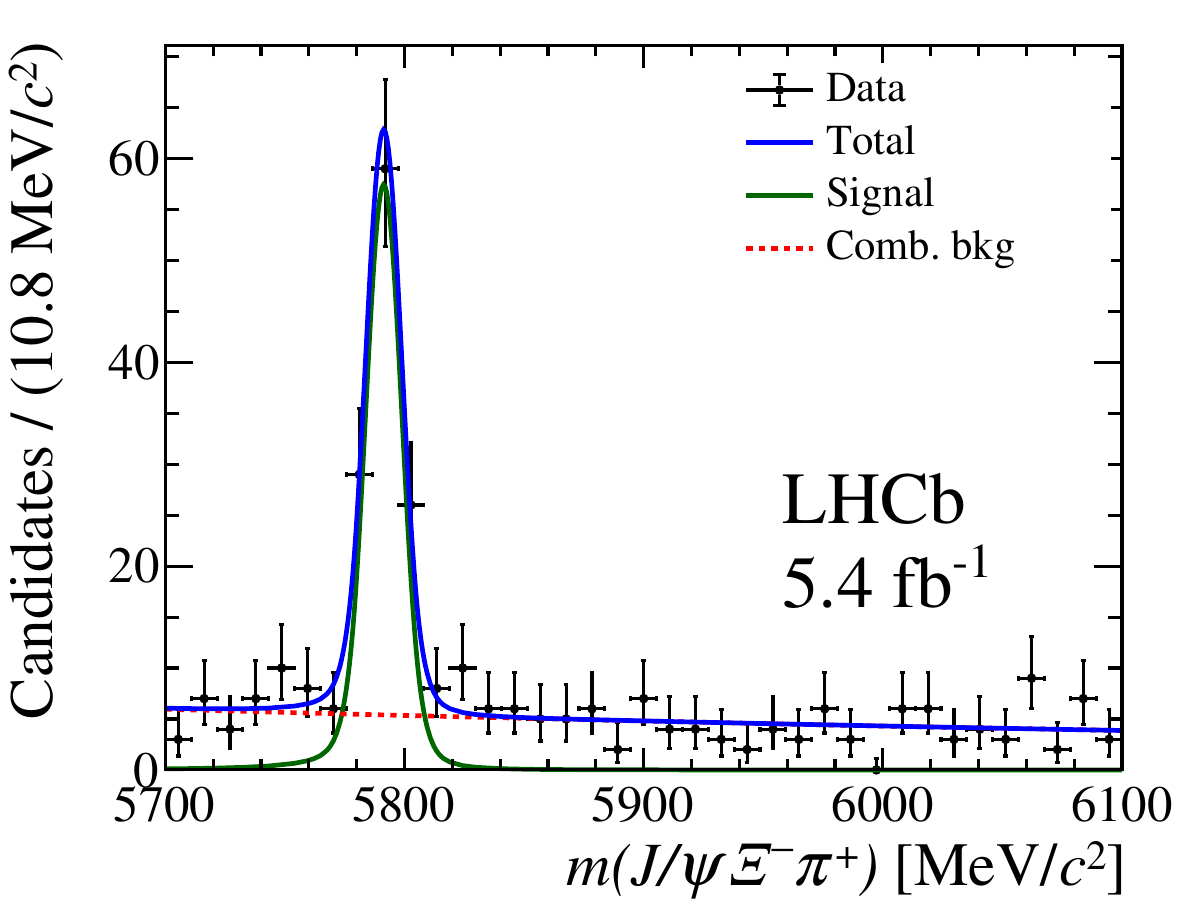}
    \includegraphics[width=0.45 \textwidth]{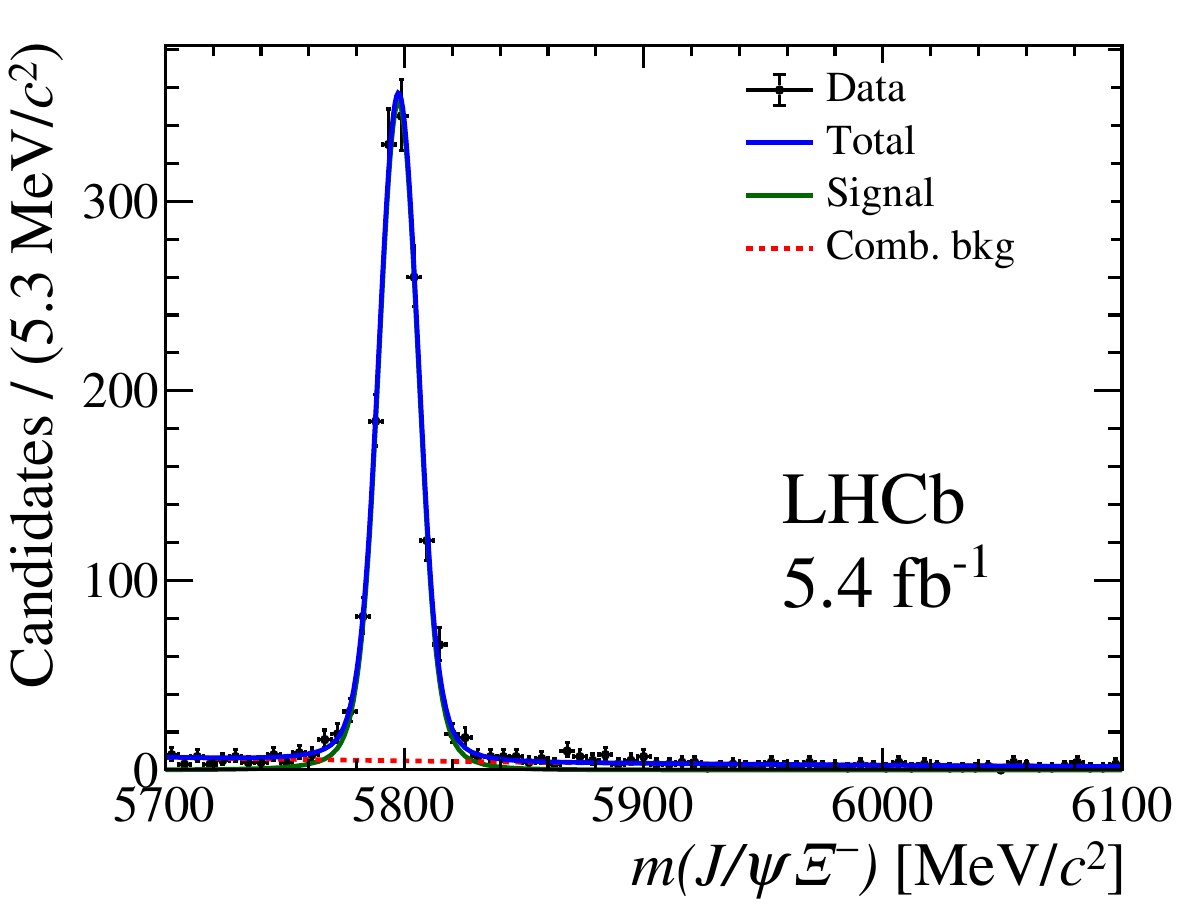}
    \caption{Invariant-mass distribution of selected (left) \XibJpsiXiPi and (right) \XibJpsiXi candidates with the result of the fit also shown. }
    \label{fig:Massfit_data_Xib}
\end{figure}

Figures~\ref{fig:Massfit_data_Lb} and \ref{fig:Massfit_data_Xib} show the result of the fits to the invariant-mass spectra for \Lb and \Xib candidates, respectively. 
The \Lb invariant-mass fits determine yields of $84 \pm 10$ \LbJpsiXiK signal and $39390 \pm 220$ \LbJpsiL normalisation decays. The \Xib invariant-mass fits determine yields of $107 \pm 12$ \XibJpsiXiPi signal and $1450 \pm 40$ \XibJpsiXi normalisation decays. Both signal modes are observed with a significance greater than $10 \sigma$ using the Wilk's theorem~\cite{Wilks:1938dza}. 

\section{Systematic uncertainties}
\label{sec:systematics}

Systematic uncertainties on the \BRRatLbText and \BRRatXibText measurements arise from several sources. A summary of the uncertainties is provided in Table~\ref{tab:systematics:summary}.
\begin{table}[!htb]
\caption{Summary of relative systematic uncertainties (in percent) for the measured ratio of branching fractions. The individual sources are described in the text. The total relative uncertainty is determined by adding the individual sources in quadrature.}
\centering 
\begin{tabular}{lcc}
\hline
Source & \BRRatLb [\%] & \BRRatXib [\%] \\
\hline
Fit model &1.9& 1.3\\ 
Size of the simulated samples & 0.9& 1.2\\ 
Corrections to simulation &  3.4 & 1.1\\ 
Alternative corrections & 1.1 & 1.2 \\
Additional corrections & 1.9 & 1.0 \\ 
Tracking efficiency & 1.5 & 1.4\\ 
Truth matching & 1.3 & 1.2 \\ 
Phase-space model & 4.5 & 3.8 \\ 
\hline
Total & 6.7 & 5.0 \\ 
\hline
\end{tabular}
\label{tab:systematics:summary} 
\end{table}

The systematic effect due to the choice of the invariant-mass fit model is assessed by means of pseudoexperiments wherein the invariant-mass distribution is generated with an alternative model and fitted using the baseline model. The alternative model replaces the two Crystal Ball functions by an Hypatia function (or vice versa for the \LbJpsiL case) for the signal. The physical backgrounds in the \LbJpsiL channel are both modelled with a single Crystal Ball function in the alternative model. 

The uncertainty on the selection efficiencies originating from the limited size of the simulated samples is propagated to the branching fraction and considered as a systematic uncertainty. 

The corrections applied to the simulation to improve the agreement with data are obtained with an uncertainty. The effect of this limited precision is assessed by varying the corrections within their uncertainty and recomputing the efficiencies 100 times. The standard deviation of the efficiency variation is assigned as a systematic uncertainty. As the corrections used for the \XibJpsiXiPi and \XibJpsiXi channels are almost the same, a good level of cancellation is obtained. On the other hand, some of the corrections applied to the \LbJpsiXiK and \LbJpsiL channels are different and, therefore they do not cancel in the ratio.

The \chisq distribution of each kinematic fit is corrected using the normalisation channels as calibration samples. While the normalisation decay topology is similar, it is not identical to the signal decay.
Alternative corrections for the simulations are derived by swapping the calibration channel, \ie using the \LbJpsiL channel for \XibJpsiXiPi corrections and the \XibJpsiXi channel for \LbJpsiXiK corrections. The variation in the efficiency is assigned as the systematic uncertainty.

After applying the baseline corrections to the simulation, some level of mismodelling remains in variables not used directly in the selection. The effect of this mismodelling is assessed by recomputing the efficiencies after correcting the additional distributions and the difference with respect to the baseline efficiency is added as a systematic uncertainty.

The signal channels feature a prompt hadron without a counterpart in the normalisation channel. An additional uncertainty on tracking efficiency of 1.4\% for pions and 1.1\% for kaons is added which accounts for hadronic interactions that do not cancel in the efficiency ratio. Additionally, the pion from the \decay{\Xim}{\Lz\pim} decay in the \LbJpsiXiK signal channel does not have a counterpart in the \LbJpsiL normalisation channel. This pion can be reconstructed as a downstream track, whose reconstruction efficiency is controlled only at the 1\% level~\cite{LHCbDownTracking}. The 1\% precision is also added as systematic uncertainty to the \Lb result.

Efficiencies are determined from simulated samples and rely on an accurate match between generated and reconstructed decays. In some cases, this so-called truth matching can fail if the fraction of hits associated between the reconstructed track and the generated particle does not reach a minimum threshold. The effect of such potential incorrect matching is evaluated by recomputing the efficiencies with an alternative method that does not rely on truth matching. Instead, the number of signal events after the selection, required to compute the efficiency, is determined by an invariant-mass fit of the \bquark-hadron spectrum, similar to the approach used for data. The difference in efficiency derived from the baseline and the alternative method is assigned as the systematic uncertainty.

Lastly, inaccuracies in the phase-space model used for signal simulation could affect the derived efficiency. The decay dynamics and angular distributions are corrected using the signal channels. The effect of the limited precision on the derived corrections is assessed by recalculating 100 times the efficiencies after varying the correction factors within their statistical uncertainty. The standard deviation is assigned as the systematic uncertainty.

The systematic uncertainties described in the section above are considered to be uncorrelated and added in quadrature.

\section{Results}
\label{sec:results}

By combining the signal and normalisation yields from the invariant-mass fits with the selection efficiency ratios as shown in Eqs.~\ref{eq:BFRatioLb} and \ref{eq:BFRatioXib}, and using the known branching fraction $\BF({\decay{\Xim}{\Lz\pim}})=0.99887\pm 0.00035$~\cite{PDG2024}, the following results are obtained:
\begin{align*}
    \BRRatLb &= (1.17 \pm 0.14 \pm 0.08)\times 10^{-2} \, , \\
    \BRRatXib &= (11.9 \pm 1.4 \pm 0.6)\times 10^{-2}\, ,
\end{align*}
where the first uncertainty is statistical, and the second is systematic.
It is possible to convert the \BRRatLbText ratio to the ${\BR(\LbJpsiXiK)/\BR(\decay{\Lb}{\psitwos\Lz})}$ ratio using the world average of ${\BR(\decay{\Lb}{\psitwos\Lz})/\BR(\LbJpsiL) = 0.508\pm 0.023}$~\cite{PDG2024} and compare it with the CMS result~\cite{Hayrapetyan:2887755}:
\begin{align*}
   \left. \frac{\BR(\LbJpsiXiK)}{\BR(\decay{\Lb}{\psitwos\Lz})}\right|_\text{CMS}&=(3.4\pm1.2)\times 10^{-2}\, , \\
    \left. \frac{\BR(\LbJpsiXiK)}{\BR(\decay{\Lb}{\psitwos\Lz})}\right|_\text{LHCb} &=(2.3\pm0.3)\times 10^{-2}\, .
\end{align*}
The \lhcb result reported in this paper is compatible  with the one obtained by CMS and improves the precision by a factor of 4.

The branching fraction of the signal channels is obtained from the measured ratios using the known branching fractions of the normalisation channels.
The values \mbox{$\BF(\LbJpsiL) = (3.36 \pm 1.11) \times 10^{-4}$} and  $\BF(\XibJpsiXi) = (5.4 \pm 2.4) \times 10^{-4}$ from Ref.~\cite{LHCb-PAPER-2021-017} are used. The signal branching fractions are determined to be:
\begin{align*}
    \BR(\LbJpsiXiK) &= (3.93 \pm 0.47 \pm 0.27 \pm 1.30)\times 10^{-6} \, ,\\
    \BR(\XibJpsiXiPi) & = (6.42 \pm 0.76 \pm 0.32 \pm 2.86)\times 10^{-5}\, ,
\end{align*}
where the first uncertainty is statistical, the second is systematic and the third is due to the external value of the normalisation branching fraction.

\begin{figure}[tb]
    \centering
    \includegraphics[width=0.45 \textwidth]{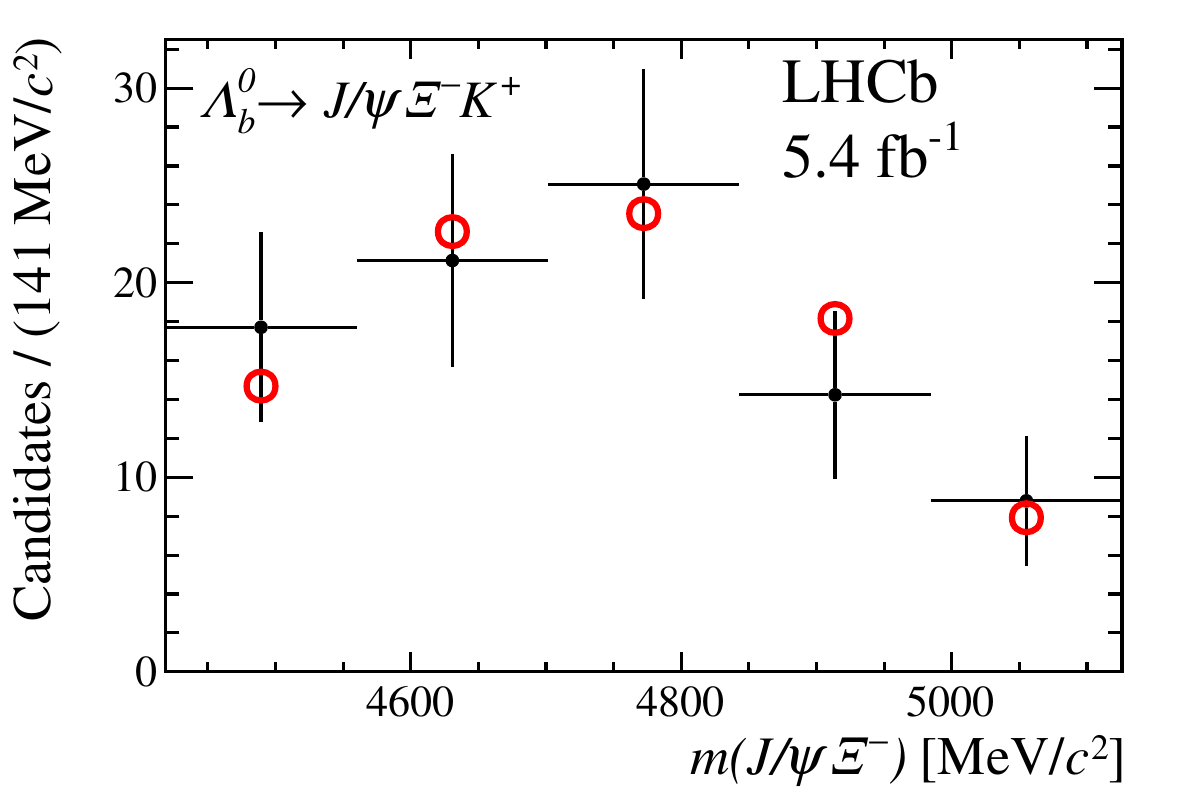}
    \includegraphics[width=0.45 \textwidth]{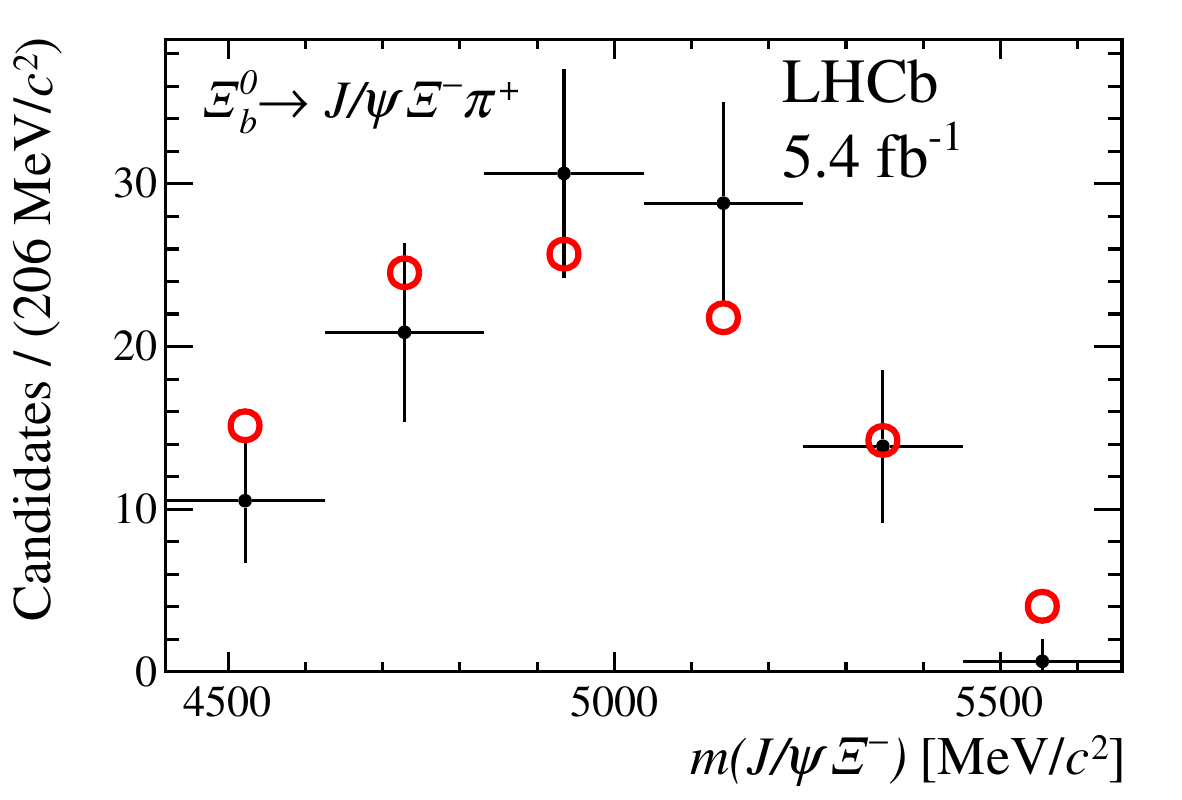}
    \includegraphics[width=0.45 \textwidth]{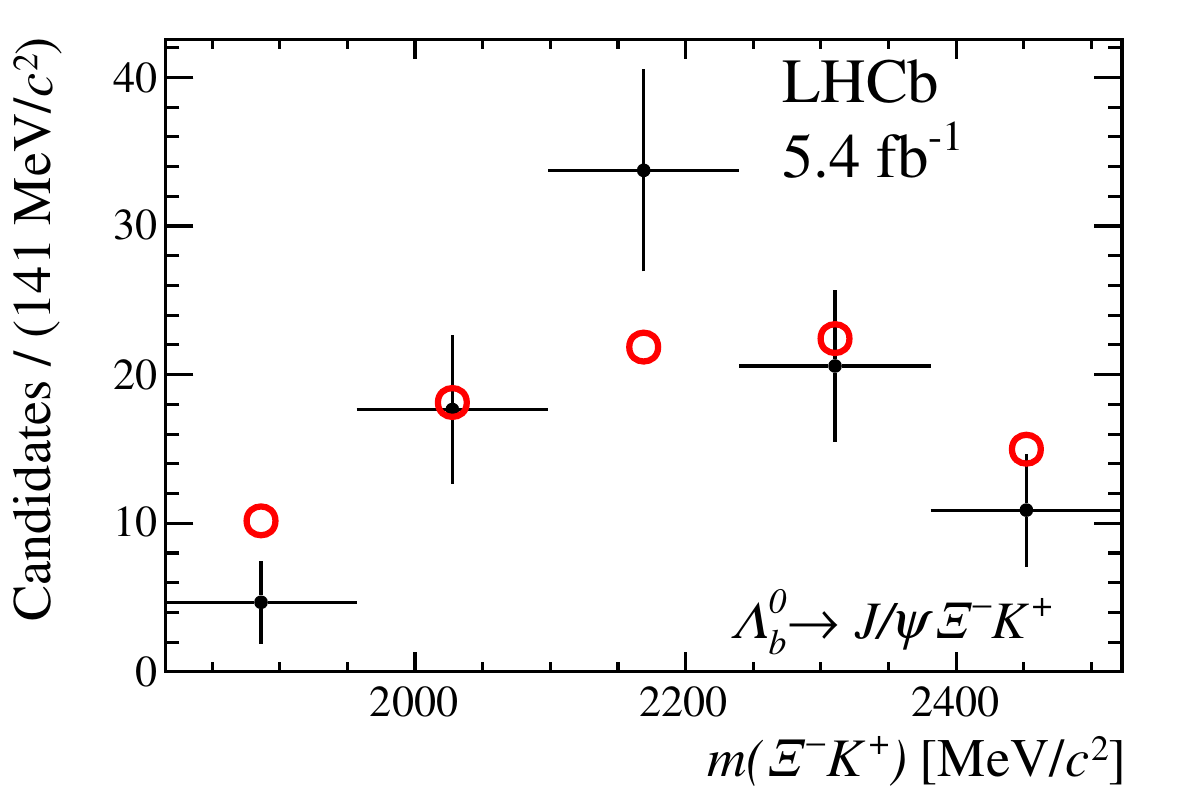}
    \includegraphics[width=0.45 \textwidth]{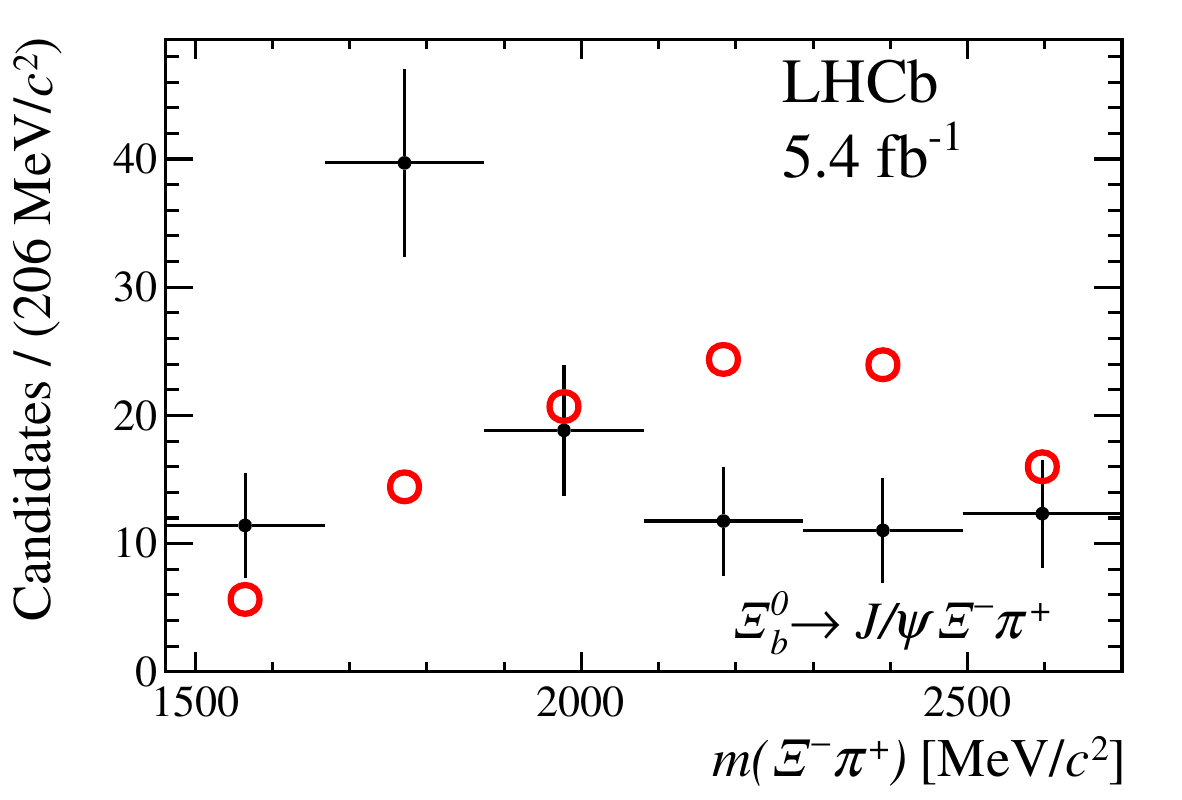}
    \includegraphics[width=0.45 \textwidth]{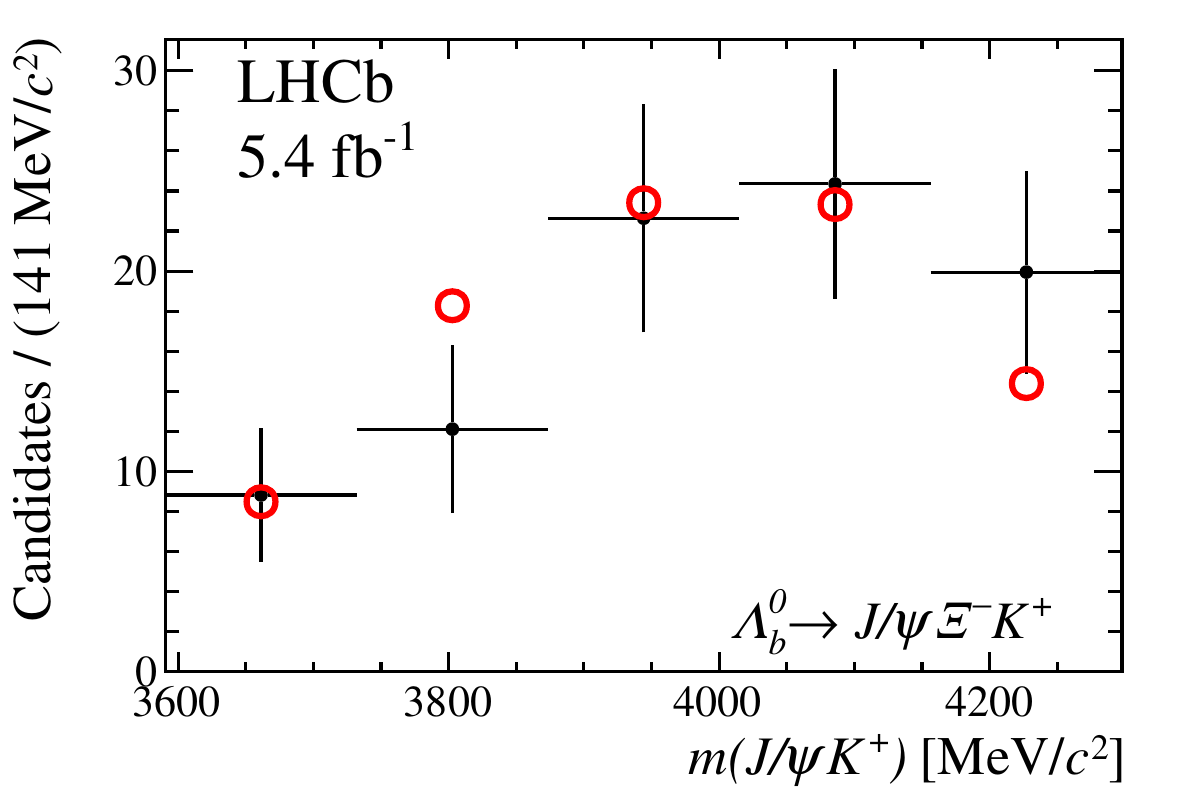}
    \includegraphics[width=0.45 \textwidth]{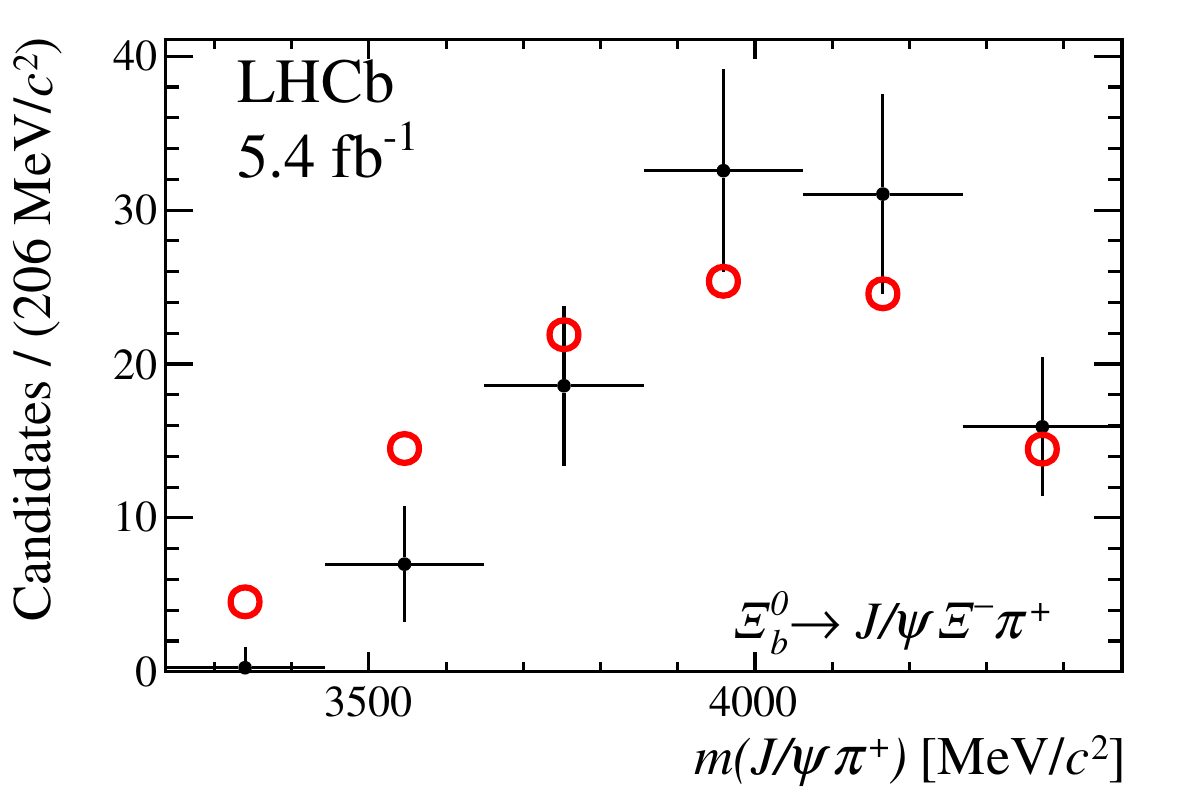}
    \caption{Invariant-mass distribution of the (top) \jpsi\Xim, (centre) \Xim\Kp or \Xim\pip and (bottom) \jpsi\Kp or \jpsi\pip systems for the (left) \LbJpsiXiK and (right) \XibJpsiXiPi decays. The (black)  background-subtracted data is compared to the (red) phase-space simulation including all weights except those for the helicity angles and two-body invariant-mass variables.}
    \label{fig:Resonance_mass_data}
\end{figure}

The background-subtracted distributions of $m(\jpsi\Xim)$, $m(\Xim\Kp)$ and $m(\Xim\pip)$ are shown in Fig.~\ref{fig:Resonance_mass_data} and are compared with the phase-space simulation. From this comparison, the only data-simulation disagreement that could point towards a resonance appears in the $m(\Xim\pip)$ range $[1670, 1850]\mevcc$ for \XibJpsiXiPi candidates. Two resonances, $\Xires(1690)$ and $\Xires(1820)$, with invariant-mass in this range are found in the literature. While the $\Xires(1690)$ resonance is observed to mainly decay to \Xim\pip,  the $\Xires(1820)$ resonance is expected to primarily decay to \Lz\Kb. Consequently, the decay channel \XibJpsiXiPi is more likely to receive contributions from the $\decay{\Xibz}{\jpsi\Xires(1690)}$ transition. However, no significant evidence supporting this hypothesis is observed in the current measurement. Larger data samples are needed to perform a quantitative study, which is left for the future.

\section{Conclusions}
In summary, a sample of \proton\proton collision data corresponding to an integrated luminosity of 5.4\invfb is used to measure the \BRRatLbText and \BRRatXibText ratios.
The \BRRatLbText result is compatible with the CMS measurement and improves its precision by a factor of four. The \XibJpsiXiPi decay is observed for the first time. The branching fraction of the \LbJpsiXiK and \XibJpsiXiPi decays are measured using the known branching fraction of the normalisation channels.
This analysis paves the way for future amplitude analysis using larger data sets that will be collected by the upgraded \lhcb experiment.

\section*{Acknowledgements}
%
%
\noindent We express our gratitude to our colleagues in the CERN
accelerator departments for the excellent performance of the LHC. We
thank the technical and administrative staff at the LHCb
institutes.
We acknowledge support from CERN and from the national agencies:
ARC and ARDC (Australia);
CAPES, CNPq, FAPERJ and FINEP (Brazil); 
MOST and NSFC (China); 
CNRS/IN2P3 (France); 
BMBF, DFG and MPG (Germany); 
INFN (Italy); 
NWO (Netherlands); 
MNiSW and NCN (Poland); 
MCID/IFA (Romania); 
MICIU and AEI (Spain);
SNSF and SER (Switzerland); 
NASU (Ukraine); 
STFC (United Kingdom); 
DOE NP and NSF (USA).
We acknowledge the computing resources that are provided by CERN, IN2P3
(France), KIT and DESY (Germany), INFN (Italy), SURF (Netherlands),
PIC (Spain), GridPP (United Kingdom), 
CSCS (Switzerland), IFIN-HH (Romania), CBPF (Brazil),
and Polish WLCG (Poland).
We are indebted to the communities behind the multiple open-source
software packages on which we depend.
Individual groups or members have received support from
Key Research Program of Frontier Sciences of CAS, CAS PIFI, CAS CCEPP, 
Fundamental Research Funds for the Central Universities, 
and Sci. \& Tech. Program of Guangzhou (China);
Minciencias (Colombia);
EPLANET, Marie Sk\l{}odowska-Curie Actions, ERC and NextGenerationEU (European Union);
A*MIDEX, ANR, IPhU and Labex P2IO, and R\'{e}gion Auvergne-Rh\^{o}ne-Alpes (France);
AvH Foundation (Germany);
ICSC (Italy); 
Severo Ochoa and Mar\'ia de Maeztu Units of Excellence, GVA, XuntaGal, GENCAT, InTalent-Inditex and Prog. ~Atracci\'on Talento CM (Spain);
SRC (Sweden);
the Leverhulme Trust, the Royal Society
 and UKRI (United Kingdom).

\clearpage




\newpage
\addcontentsline{toc}{section}{References}
\bibliographystyle{LHCb}
\bibliography{main,standard,LHCb-PAPER,LHCb-CONF,LHCb-DP,LHCb-TDR}

 
\newpage
\centerline
{\large\bf LHCb collaboration}
\begin
{flushleft}
\small
R.~Aaij$^{38}$\lhcborcid{0000-0003-0533-1952},
A.S.W.~Abdelmotteleb$^{57}$\lhcborcid{0000-0001-7905-0542},
C.~Abellan~Beteta$^{51}$,
F.~Abudin{\'e}n$^{57}$\lhcborcid{0000-0002-6737-3528},
T.~Ackernley$^{61}$\lhcborcid{0000-0002-5951-3498},
A. A. ~Adefisoye$^{69}$\lhcborcid{0000-0003-2448-1550},
B.~Adeva$^{47}$\lhcborcid{0000-0001-9756-3712},
M.~Adinolfi$^{55}$\lhcborcid{0000-0002-1326-1264},
P.~Adlarson$^{82}$\lhcborcid{0000-0001-6280-3851},
C.~Agapopoulou$^{14}$\lhcborcid{0000-0002-2368-0147}
% ...

{\footnotesize \it

% ...
$^{14}$Universit{\'e} Paris-Saclay, CNRS/IN2P3, IJCLab, Orsay, France\\
% ...
$^{38}$Nikhef National Institute for Subatomic Physics, Amsterdam, Netherlands\\
% ...
$^{47}$Instituto Galego de F{\'\i}sica de Altas Enerx{\'\i}as (IGFAE), Universidade de Santiago de Compostela, Santiago de Compostela, Spain\\
% ...
$^{51}$Physik-Institut, Universit{\"a}t Z{\"u}rich, Z{\"u}rich, Switzerland\\
% ...
$^{55}$H.H. Wills Physics Laboratory, University of Bristol, Bristol, United Kingdom\\
% ...
$^{57}$Department of Physics, University of Warwick, Coventry, United Kingdom\\
% ...
$^{60}$School of Physics and Astronomy, University of Glasgow, Glasgow, United Kingdom\\
$^{61}$Oliver Lodge Laboratory, University of Liverpool, Liverpool, United Kingdom\\
% ...
$^{69}$Syracuse University, Syracuse, NY, United States\\
% ...
$^{82}$Department of Physics and Astronomy, Uppsala University, Uppsala, Sweden, associated to $^{60}$\\
% ...
}
\end{flushleft}

\end{document}